\documentclass[a4paper,10pt,twoside]{cpc-hepnp}

\usepackage{multicol}
\usepackage{graphicx}
\usepackage{booktabs}
\usepackage{amssymb,bm,mathrsfs,bbm,amscd}
\usepackage[tbtags]{amsmath}
\usepackage{lastpage}
\usepackage{epstopdf}

\begin{document}

\fancyhead[c]{\small Submitted to Chinese Physics C} \fancyfoot[C]{\small 010201-\thepage}

\title{Some new progress on the light absorption properties of
linear alkyl benzene solvent\thanks{Supported by National 973 Project Foundation of the Ministry of Science and Technology of China (Contract No. 2013CB834300)}}

\author{%
      YU Guang-You$^{1}$
\quad CAO De-Wen$^{1}$
\quad HUANG Ai-Zhong$^{2}$\\
\quad Yu Lei$^{3}$
\quad LOH Chang-Wei$^{1}$
\quad WANG Wen-Wen$^{1}$\\
\quad QIAN Zhi-Qiang$^{1}$
\quad YANG Hai-Bo$^{1}$
\quad HUANG Huang$^{1}$\\
\quad XU Zong-Qiang$^{1}$
\quad ZHU Xue-Yuan$^{1}$
\quad XU Bin$^{1}$
\quad Qi Ming$^{1;1)}$\email{qming@nju.edu.cn}%
}
\maketitle

\address{%
$^1$ National Laboratory of Solid State Microstructures and School of physics, Nanjing University,\\ Nanjing, 210093, People's Republic of China\\

$^2$ Jinling Petrochemical Corporation Ltd, Nanjing, 210000, People's Republic of China\\

$^3$ National 863 Program New Material MO Precursors R\&D Center, Nanjing University,\\ Nanjing, 210093, People's Republic of China\\

}

\begin{abstract}
Linear alkyl benzene (LAB) will be used as the solvent of a liquid scintillator mixture for the JUNO antineutrino experiment in the near future. Its light absorption property should therefore be understood prior to its effective use in the experiment. Attenuation length measurements at a light wavelength of 430 nm have been performed on samples of LAB prepared for the purpose of the JUNO experiment. Inorganic impurities in LAB have also been studied for their possibilities of light absorption in our wavelength of interest. In view of a tentative plan by the JUNO collaboration to utilize neutron capture with hydrogen in the detector, we have also presented in this work, a study on the carbon-hydrogen ratio and the relationship thereof with the attenuation length of the samples.
\end{abstract}



\begin{multicols}{2}

\section{Introduction}

The Jiangmen Underground Neutrino Observatory (JUNO) is a multi-purpose experiment scheduled to start its first data-taking in 2020 with an aim to study the neutrino mass hierarchy \cite{lab1,lab2,lab3,lab4} for any physics beyond the Standard Model. As antineutrinos interact weakly with other particles, the JUNO detector needs to be large in size and consequently needs about 20 kilotons of LAB as a liquid scintillator (LS) solvent.

The sheer size of the detector poses also a challenge in that the liquid scintillator to be used in the detector for the purpose of reacting with incoming antineutrinos is required to have high optical transparency and an attenuation length comparable to the size of the detector itself. Otherwise, photons emitted through a series of interactions between the antineutrinos and LS will be absorbed by the LS itself before reaching the photomultiplier tubes (PMTs) outside of the JUNO detector. As the inner sphere of the JUNO central detector having a diameter of about 34 m will contain the liquid scintillator comprising of a linear alkyl benzene solvent with PPO and bis-MSB as the solutes, the attenuation length of the LAB solvent itself should be comparable to the said diameter, in order to prepare a liquid scintillator with minimal absorption of the emitted photons. As a note, the LAB quality in use currently contained in the Daya Bay detector \cite{lab5} has an attenuation length about 10 m or slightly larger, which is still less than the requirement of JUNO.

It is imperative therefore, that effort on producing LAB with better light absorption properties continues to be done. To date, Johnny Goett et al. \cite{lab6} and Yayun Ding et al. \cite{lab7} have studied on the attenuation length of LAB purified and prepared in the laboratory, with the former obtaining an attenuation length of 28.6 m and the latter obtaining 26 m from their respective samples. Looking forward, these results indicate that the aim to prepare LAB with an attenuation length of at least 30 m and comparable to the size of the JUNO detector is an achievable feat in the near future. In this work, we present a study on the light absorption properties of new LAB samples obtained through a scalable LAB manufacturing process from Jinling Petrochemical Corporation Ltd (hereafter known as "Jinling") which can be feasibly used to do a mass production of LAB as required by the JUNO detector in the coming future. The attenuation length of the samples has been measured using an updated setup with improved techniques that differ from that used for the purpose of the Daya Bay experiment \cite{lab8}. The largest attenuation length as measured from our samples was found to be 26.8 $\pm$ 0.4 m.

In addition, we have used an ICP-MS method to study the inorganic impurities in LAB in contrast with the GC-MS and LC-MS methods to study the organic impurities in \cite{lab8} and \cite{lab9}. At present, we found the concentration of the inorganic impurities to be so low that any significant influence thereof on the attenuation length of LAB can be ignored provisionally. The C/H ratio of LAB and comparison between the samples based on the ratios and attenuation lengths has also been investigated. It is found that the C/H ratio has a non-trivial relationship with the attenuation length of the LAB samples.

The results of this work and the setup described herein will be useful for future studies in the preparation of LAB and finally LS with requirements for the light absorption properties as stringent as that from JUNO.

\section{LAB samples}

The LAB samples used in this work tagged as NJ25\#, NJ26\#, NJ28\#, NJ29\#, NJ30\#, NJ32\# and NJ33\#, were provided by Jinling. Specifically, NJ29\#, NJ30\#, NJ32\# and NJ33\# were prepared with some improved techniques that can help form a basis for large-scale manufacturing of LAB with large attenuation length in Jinling. NJ28\# is a sample taken from the LS used in the Daya Bay detector, doped with PPO, bis-MSB and gadolinium (Gd) inside. A sample of LAB without the addition of PPO, bis-MSB and Gd, tagged as NJ22\#, originally prepared for the Daya Bay experiment, will be used as a reference in this work.

\section{Attenuation Length of LAB}

\subsection{Definition}

The attenuation length $L_{\lambda}$ is defined to be the distance in a material where the intensity of an incident light with wavelength $\lambda$ is reduced by 1/e. This can be expressed as
\begin{equation}
I=I_{0}e^{-\frac{x}{L_{\lambda}}},
\end{equation}
where $I$ is the light intensity after it passes through the material, which in this work will be the LAB liquid, for a total path length $x$, and $I_{0}$ is the initial light intensity.
\subsection{Experimental Setup}

In this work, LAB is filled into a 1.2 m steel tube with Teflon-covered inner wall. The photomultiplier tube is placed at the bottom of the tube, and an LED with a mean wavelength of 430 nm is placed at the top of the tube acting as the light source. This is similar to that performed in \cite{lab8,lab10,lab11}.

\end{multicols}{}

\begin{center}
\includegraphics[width=12cm]{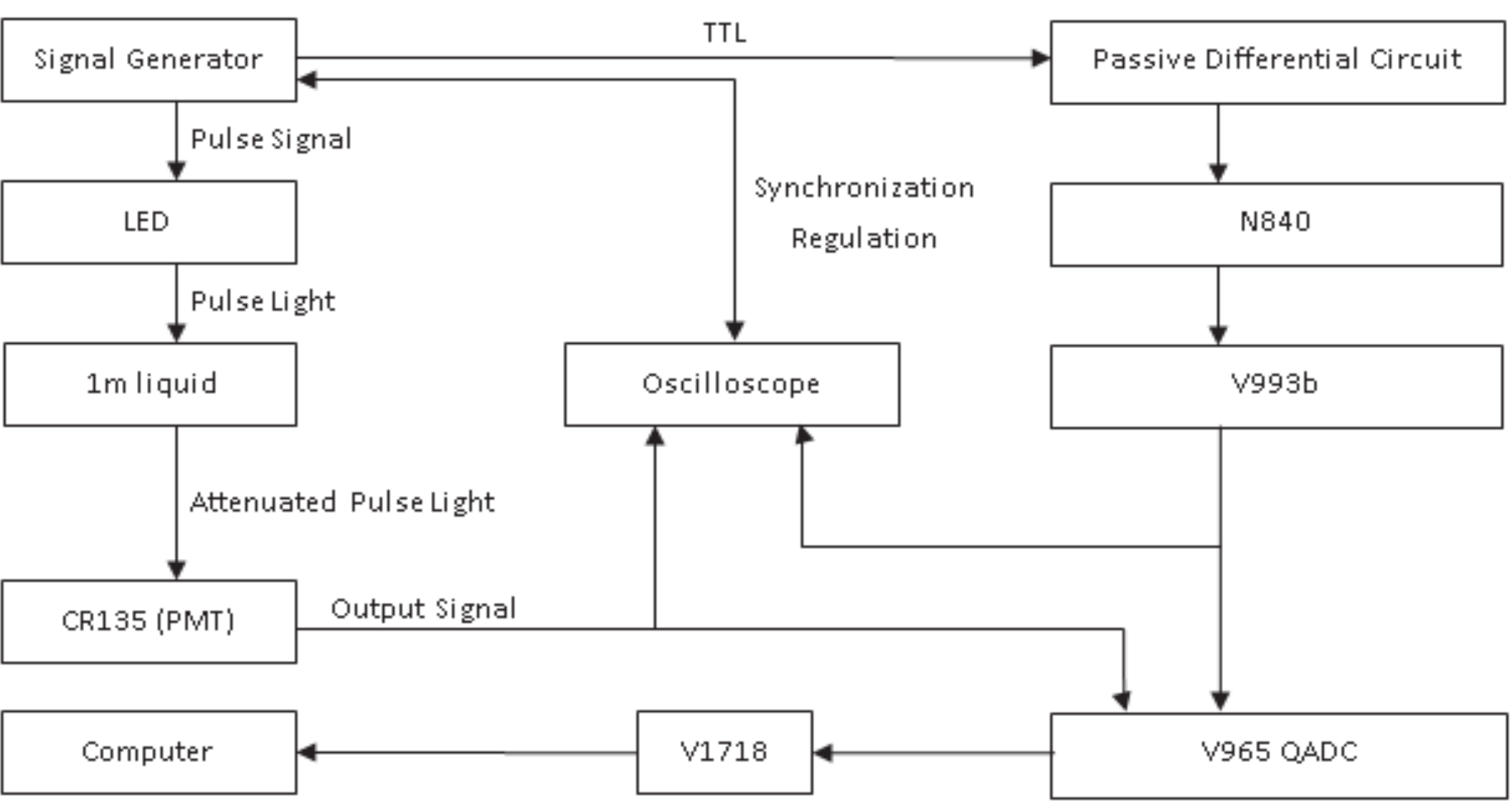}
\figcaption{\label{fig1}Schematic diagram of the DAQ system.}
\end{center}

\begin{multicols}{2}

Using a novel DAQ system with reference to \cite{lab12} as shown schematically in Fig. 1, two signals with the same frequency are generated by the signal generator. One of the signals is to drive the LED to flash. The LED light pulse then travels down the tube containing LAB to the PMT at the bottom of the tube. The generated electronic signal from the PMT is then transmitted to V965, a Dual Range Charge to Amplitude Converter, to convert the PMT signal based on the strength of the attenuated light to an $ADC$ channel value. The other signal from the signal generator is transmitted to V965 via a passive differential circuit, N840 and V993 to form a gate signal. The oscilloscope is used for the synchronization regulation between the PMT signal and the gate signal. Using the $ADC$ channel values acquired from the PMT signal, Eq. (1) can be modified to be
\begin{equation}
ADC_{x}=\alpha ADC_{0}e^{-\frac{x}{L_{\lambda}}},
\end{equation}
where $ADC_{x}$ is the mean $ADC$ channel value when the liquid height is $x$, $ADC_{0}$ is the $ADC$ channel value when the liquid height is 0.1 m, and $\alpha$ is simply a multiplicative factor due to the fact that we have substituted the initial light intensity-related $ADC$ with an $x=0.1$ m light intensity-related $ADC$. In performing the experiment, we have measured the mean $ADC$ channel value at every 0.1 m interval beginning from about 1 m in a decreasing manner, i.e. $|\Delta x|=0.1$ m.

For the purpose of finding the optimal PMT applied high voltage, we have performed a single photon experiment using the same setup and DAQ system as aforementioned. The corresponding signal from a single photon is shown in Fig. 2. We found that performing the experiment under a PMT high voltage of 1100 V led to the clearest single photon peak. Such attainability shows that our setup and the DAQ system have a high sensitivity. The data acquisition and transmission rate is determined experimentally. We found that, after a large number of trials, the frequency of the signal generator at 800 Hz was the most optimal. As such, the PMT high voltage was set to 1100 V, and the signal generator frequency was set to 800 Hz throughout the entire experiment for the attenuation length measurements.

\begin{center}
\includegraphics[width=8cm]{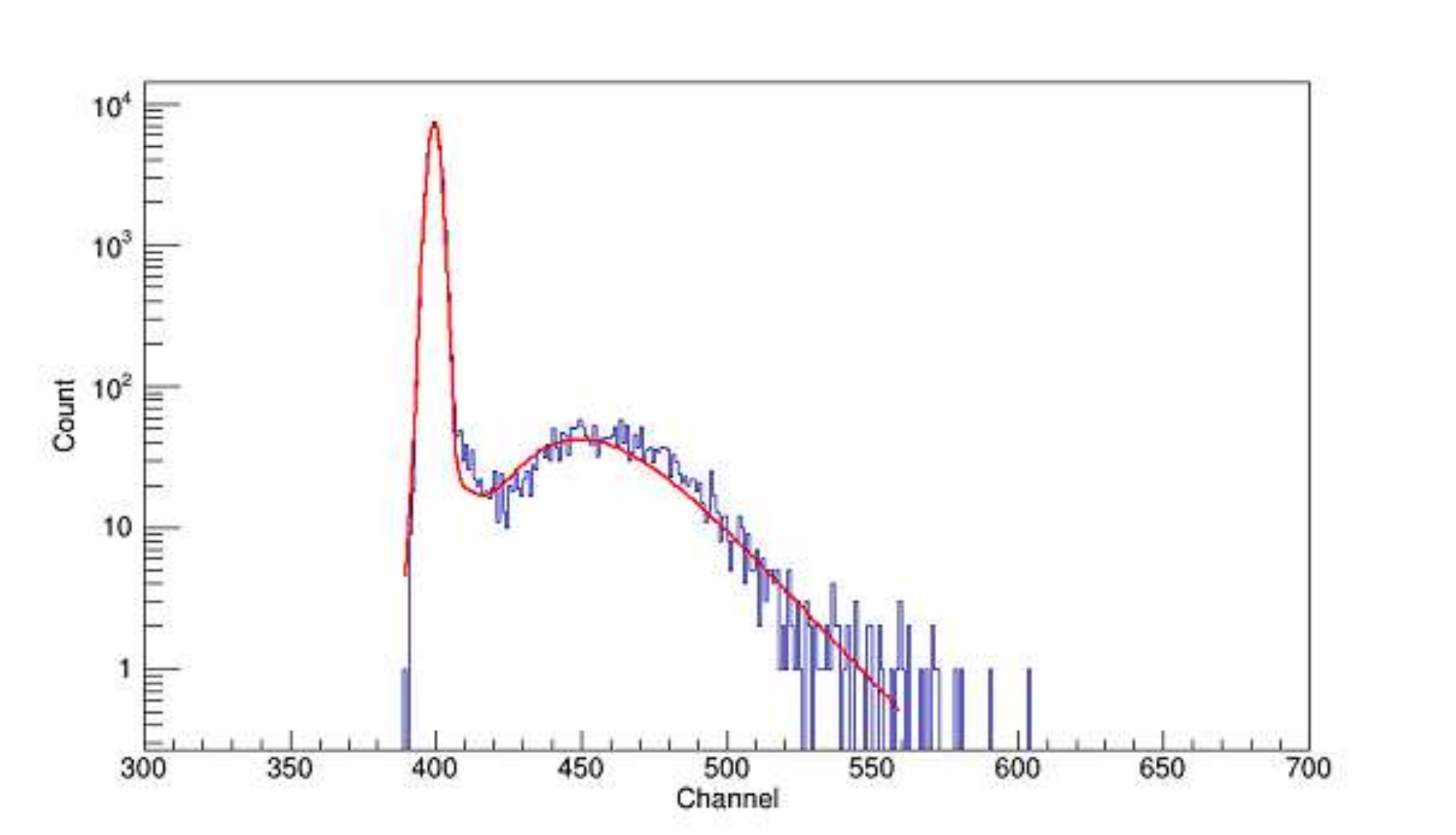}
\figcaption{\label{fig2}Signal from a single photon. The x-axis
corresponds to the $ADC$ channel values. Each channel corresponds to 200 fC.}
\end{center}

\end{multicols}{}

\begin{center}
\includegraphics[width=12cm]{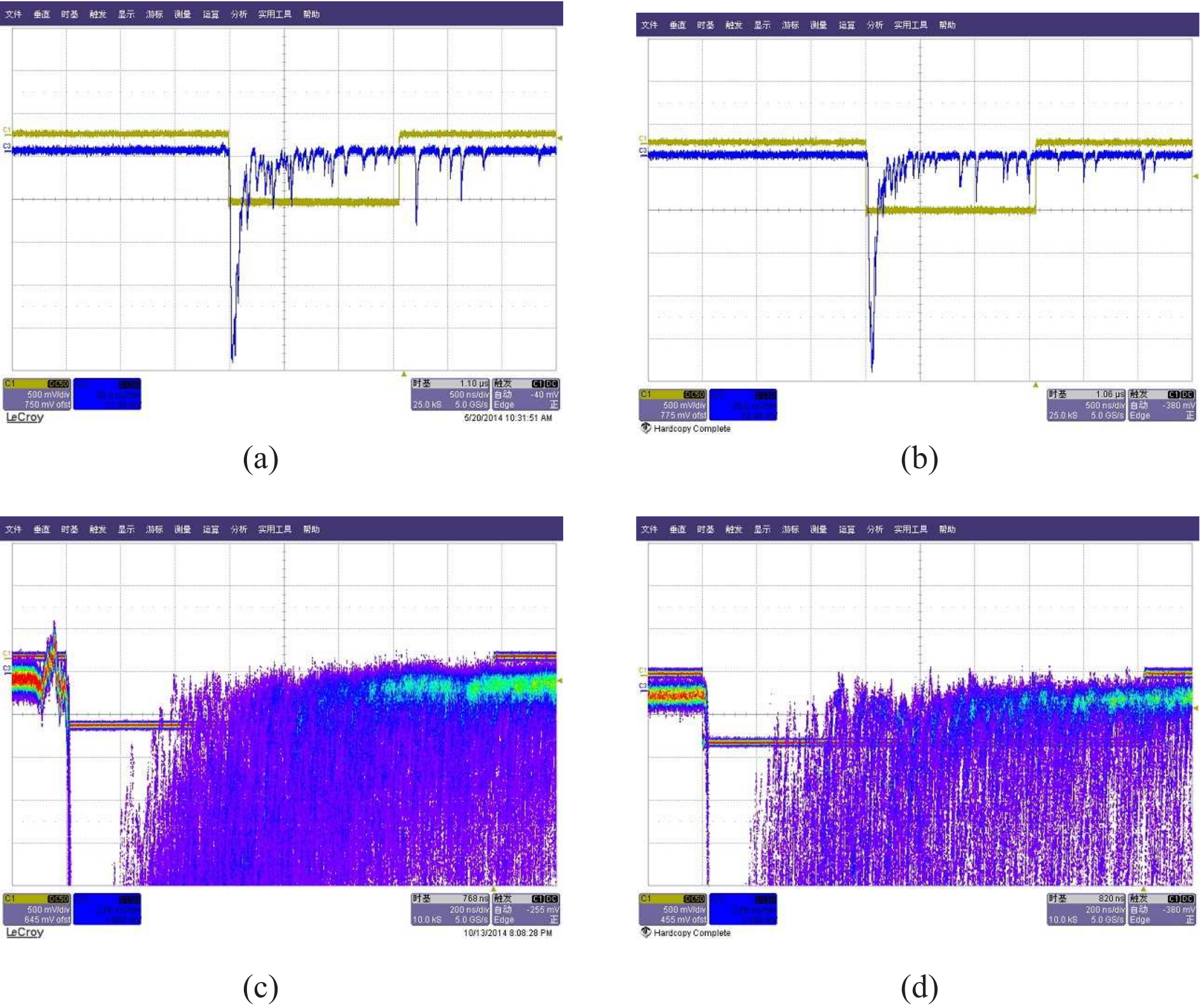}
\figcaption{\label{fig3}(a) and (c): A scope view of the PMT signal before placing the experimental setup in the clean dark room. (b)  and (d): A scope view of the PMT signal after placing the experimental setup in the clean dark room.}
\end{center}

\begin{multicols}{2}

\subsection{Improvement and upgrade}

To reduce the effect of stray light, dust and fluctuations in the ambient temperature to the LED, pulse generator, PMT and electronic readout system, a clean dark room has been constructed and is placed in our lab with double pane windows. The walls of the room formed a Faraday cage with good ground connection to reduce electromagnetic disturbances to the experiment. Likewise, the PMT signal transmission cables are of shielded types to also reduce electromagnetic disturbances. Fig. 3(a) and Fig. 3(c) show a PMT signal without having the experimental setup in the clean dark room in comparison with a signal with the clean dark room as shown in Fig. 3(b) and Fig. 3(d). The usage of our improved setup in the clean dark room has successfully reduced the number and magnitude of the after-pulses as shown in Fig. 3(b) compared to Fig. 3(a). In addition, all pre-pulses have disappeared as can clearly be seen from Fig. 3(d) compared to Fig. 3(c) which was obtained prior to using the clean dark room.

As the PMT signal output has a negative correlation with the ambient temperature, the fluctuation in the latter can lead to a large uncertainty in the readout and the measurement of the attenuation length. The relation between the ambient temperature and the $ADC$ value of the PMT signal is shown in Fig. 4 and 5. The results in Fig. 4, taken over a period of four days, were obtained when the experimental setup was not in the dark room. The results in Fig. 5, taken over more than a 10-hour period, were obtained when the experimental setup was placed in the dark room. Fig. 4 shows that the temperature can fluctuate within 5 $^{\circ}$C leading to a 7.5 \% fluctuation in the PMT signal output. Fig. 5 shows that the temperature fluctuation can be stabilized to below 0.5 $^{\circ}$C leading to a mere 0.2 \% fluctuation in the PMT signal output. This can be observed from the 400 minute mark to the 680 minute mark in Fig. 5. This stable period was more than sufficient for us to perform the necessary measurements on a LAB sample. Using $\Delta ADC/ADC =|\Delta x|/L_{\lambda}$, and that $|\Delta x|=0.1$ m and $\Delta ADC/ADC =0.2 \%$, an attenuation length $L_{\lambda}$   of larger than 30 m can be measured in principle.

\begin{center}
\includegraphics[width=8cm]{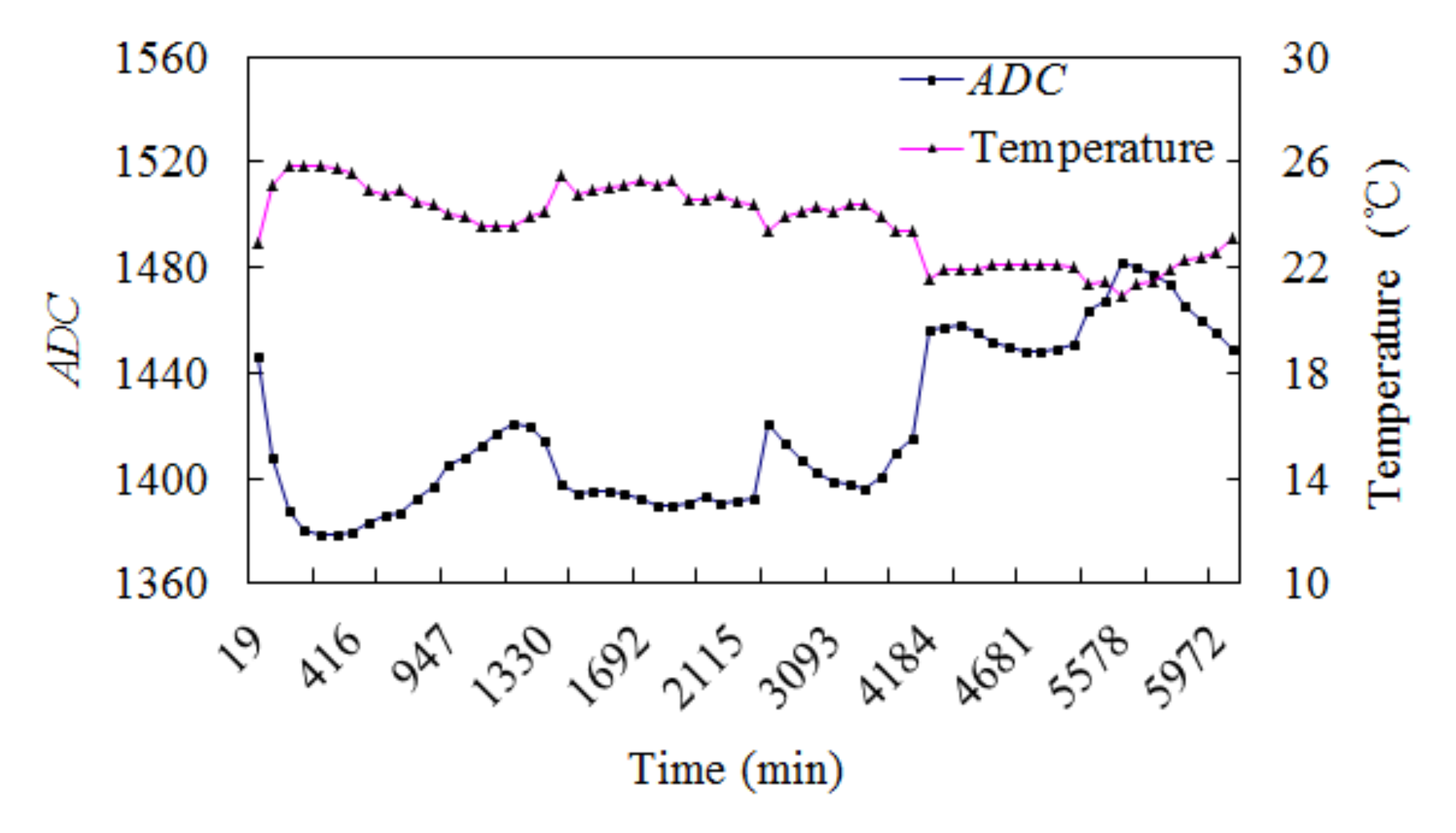}
\figcaption{\label{fig4}Change in the $ADC$ value and the ambient temperature when the setup is not placed in a dark room.}
\end{center}

\begin{center}
\includegraphics[width=8cm]{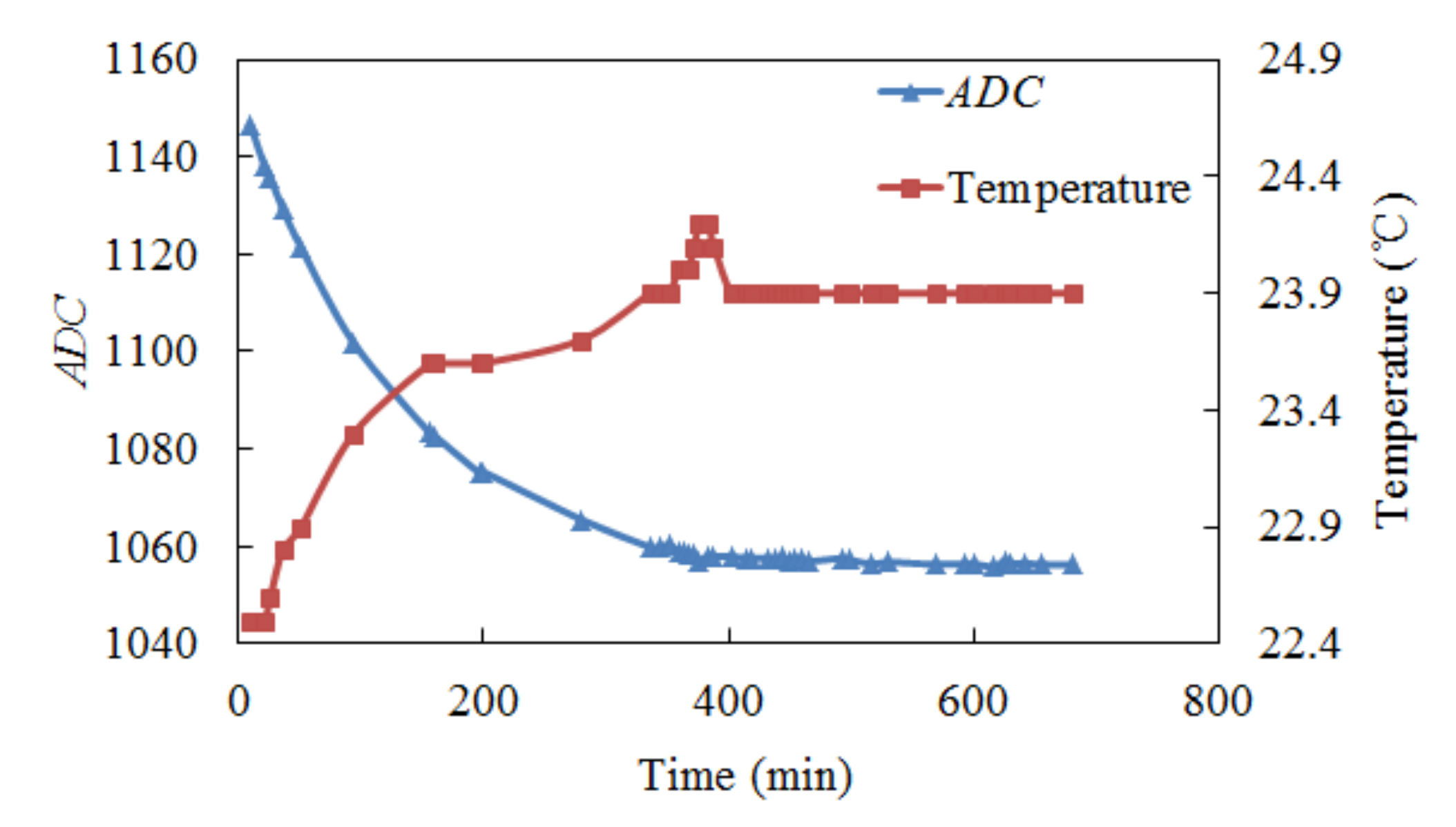}
\figcaption{\label{fig5}Change in the $ADC$ value and the ambient temperature when the setup is in a dark room.}
\end{center}

\section{Results}

\subsection{Attenuation length measurements of LAB}

Fig. 6 shows the measurement results of NJ33\# based on the usage of different light intensities. The first and second trial has been performed with the same light intensity obtaining similar attenuation length within the uncertainty showing the stability of the experimental setup.

\begin{center}
\includegraphics[width=8cm]{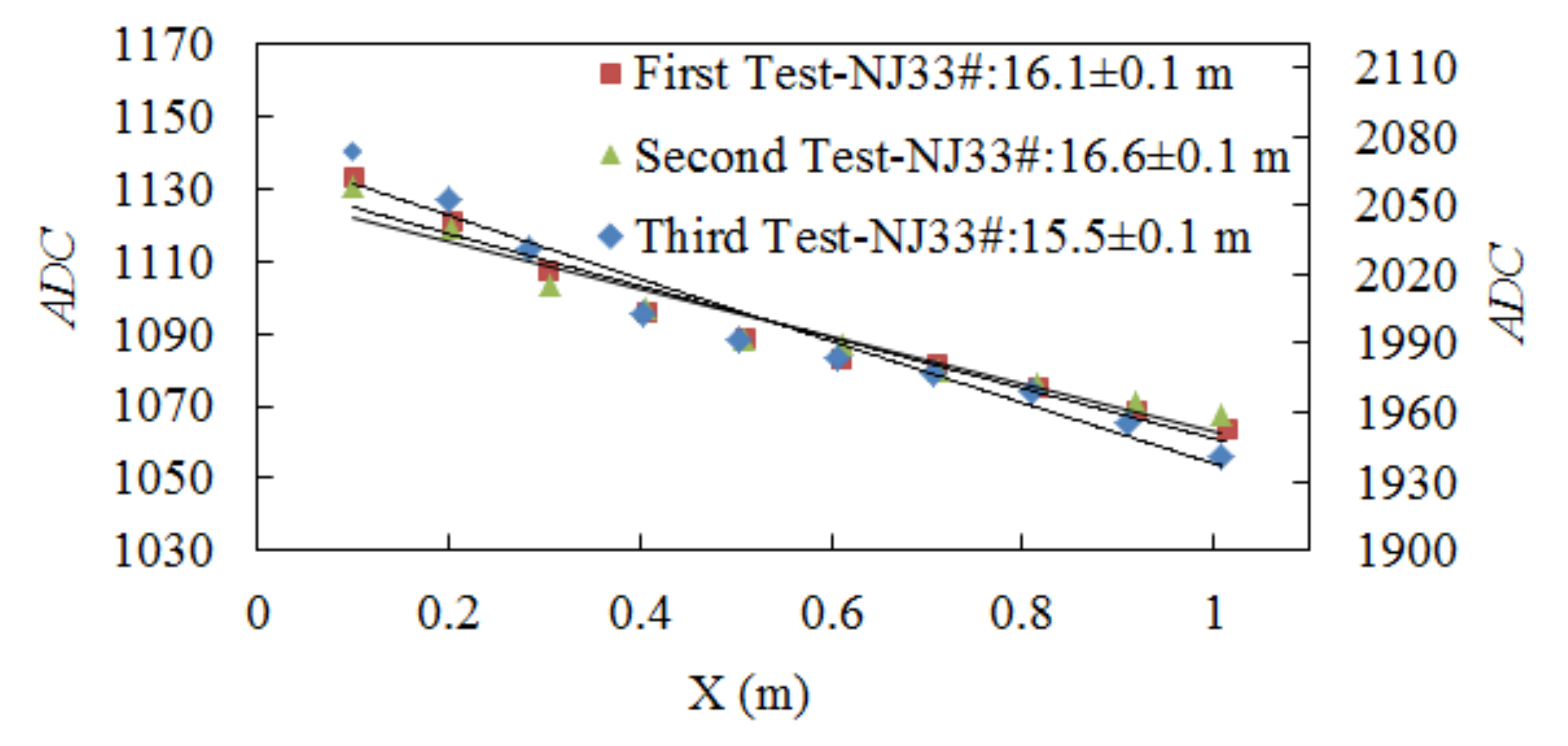}
\figcaption{\label{fig6}Three trials of attenuation length measurements have been performed with NJ33\#. The first and second trial has been performed with the same light intensity, while the third trial has been performed with a 50 \% larger light intensity compared to the first and second trial. Included is the attenuation length obtained for each trial from an exponential fit to the data.}
\end{center}

Fig. 7 shows the measurement results of five LAB samples. Their corresponding attenuation lengths as obtained from the fits to the data are shown in Table 1. An investigation has been made as to whether the method of reducing the liquid height of LAB has any impact on the measurements of the attenuated light intensity. Fig. 8 shows the results for NJ32\# when the liquid height is reduced in steps of 0.02 m until the next desired liquid height to be measured for its corresponding attenuated light intensity.

\begin{center}
\includegraphics[width=8cm]{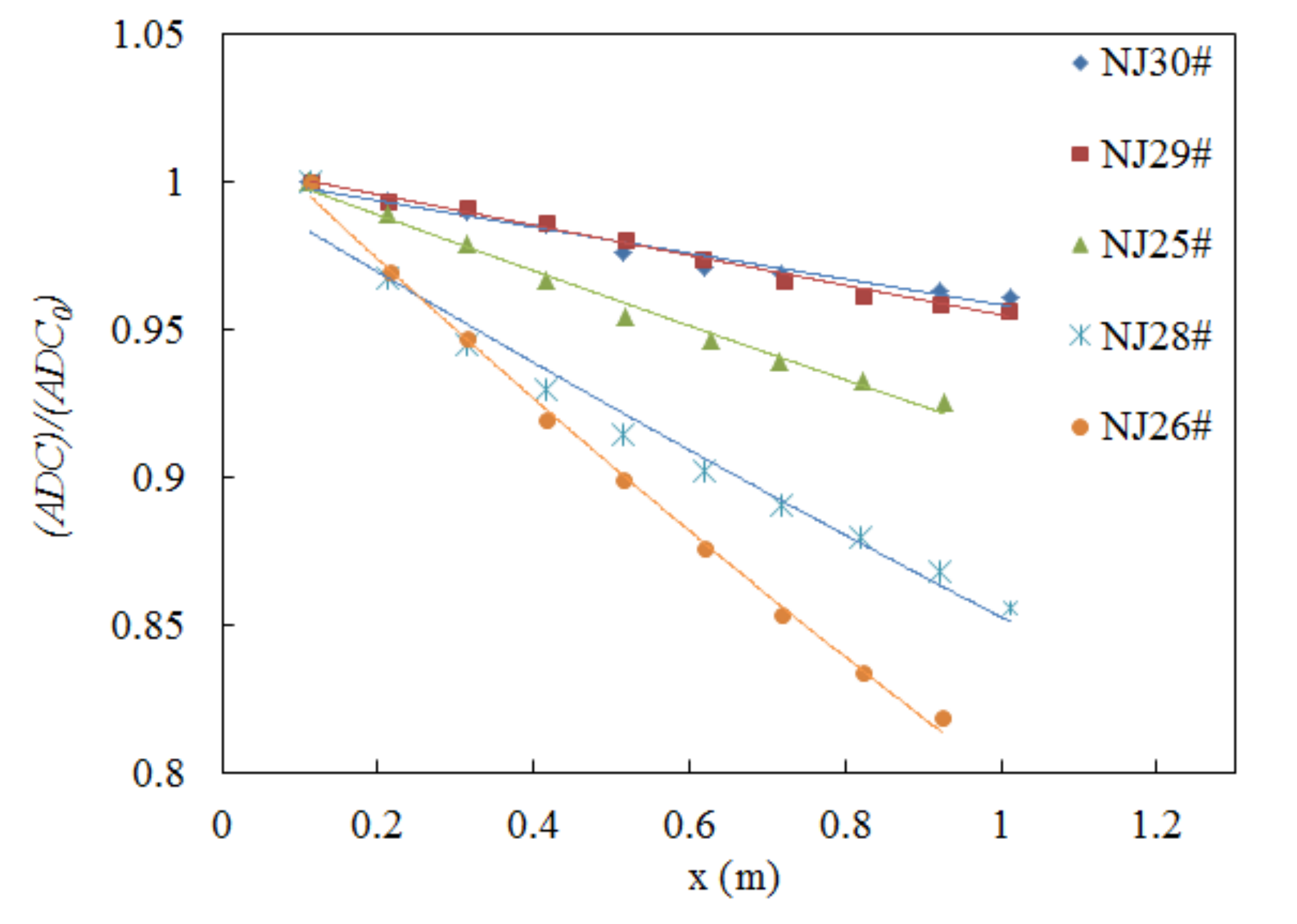}
\figcaption{\label{fig7}$ADC$/$ADC_{0}$ of NJ25\#, NJ26\#, NJ28\#, NJ29\# and NJ30\# for various LAB liquid height x as measured with a low light intensity.}
\end{center}

\begin{center}
\tabcaption{ \label{tab1}Attenuation length of LAB samples.}
\footnotesize
\begin{tabular*}{80mm}{c@{\extracolsep{\fill}}ccc}
\toprule LAB samples & Attenuation length (m)\\
\hline
NJ22\# & 11.2 $\pm$ 0.5  \\
NJ25\# & 10.2 $\pm$ 0.1  \\
NJ26\# & 4.02 $\pm$ 0.01  \\
NJ28\# & 6.19 $\pm$ 0.02 \\
NJ29\# & 19.2 $\pm$ 0.2 \\
NJ30\# & 22.1 $\pm$ 0.3 \\
\bottomrule
\end{tabular*}
\end{center}

\begin{center}
\includegraphics[width=8cm]{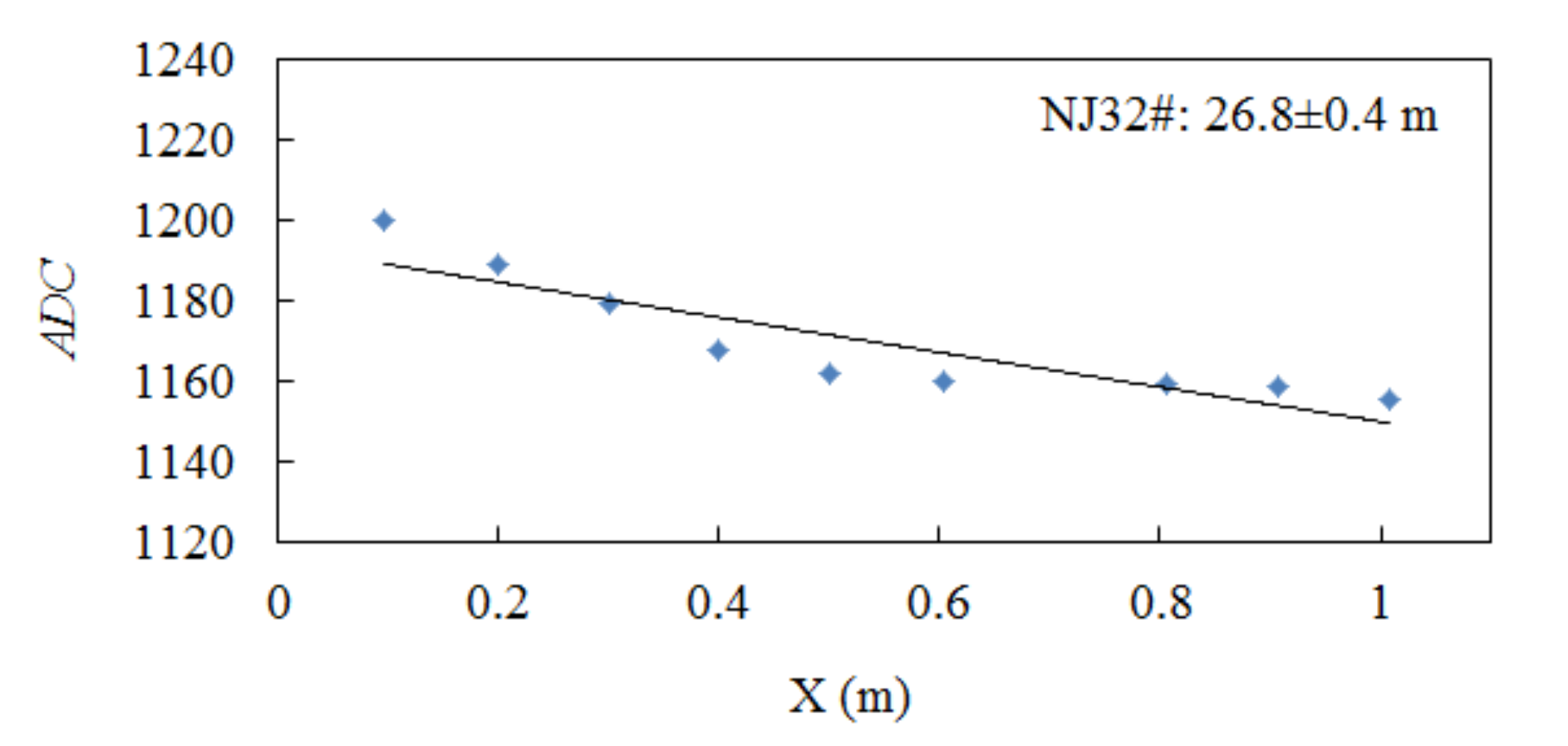}
\figcaption{\label{fig8}$ADC$ for NJ32\# for various LAB liquid height x, wherein the reduction in x is done in steps of 0.02 m until the next desired liquid height to be measured for its corresponding attenuated light intensity. The attenuation length as obtained from the exponential fit is 26.8 m.}
\end{center}

\subsection{Inorganic impurities in LAB}

Clathrates can be formed when inorganic impurities are dissolved in LAB, which can produce ligand centered (LC) $ \pi \rightarrow \pi^{*}$, metal to ligand charge-transfer (MLCT) and metal to metal charge-transfer (MMCT) excitations therein. These excitations are in the light-absorption ranges with a wavelength of 350 nm to 550 nm \cite{lab13,lab14}, which can possibly reducing the attenuation length of our LAB samples in this wavelength range. Using the ICP-MS (Thermo X series II) technique, we have studied several impurities in LAB, including Li, B, Na, Mg, Al, K, Ca, Ti, Cr, Mn, Fe, Co, Ni, Cu, Zn, As, Sr, Ag, Cd, Sn, Ba and Pb. Table 2 shows the impurity contents by elements in several samples. Elements studied under the ICP-MS technique but are not listed in Table 2 were found to have concentrations below the baseline of the technique.

\begin{center}
\tabcaption{ \label{tab1}Concentration of inorganic impurities in the LAB samples.}
\footnotesize
\begin{tabular*}{80mm}{c@{\extracolsep{\fill}}ccc}
\toprule    & NJ29\# & NJ28\# & NJ26\#\\
\hline
Na/ppb & $<0.1$ & 406.1 & $<0.1$ \\
Fe/ppb & $<0.1$ & $<0.1$ & $<0.1$ \\
Cu/ppb & $<0.01$ & $<0.01$ & $<0.01$ \\
Zn/ppb & $<0.01$ & $<0.11$ & $<0.01$ \\
Ru/ppb & $<0.01$ & $<0.01$ & $<0.01$ \\
Ag/ppb & $<0.01$ & 0.02 & $<0.01$ \\
Os/ppb & $<0.01$ & 0.23 & $<0.01$ \\
Au/ppb & $<0.01$ & 0.04 & $<0.01$ \\
\bottomrule
\end{tabular*}
\end{center}

As a reference, Gd has a concentration of 200 ppm in the LS liquid used in the Daya Bay experiment and was found to only reduce the attenuation length of its LS sample from 15.99 m to 15.10 m \cite{lab15}.

Based on the aforementioned and Table 2, we find that the concentration of the inorganic impurities in our LAB samples is so low, and as such, any possibility of clathrates that can reduce the attenuation length of LAB is provisionally negligible at the moment.

\subsection{C/H ratio of LAB}

Gd has been used in past detectors including Daya Bay for enhancement of neutron capture \cite{lab16}. In view of a tentative plan by JUNO to use an LS liquid without the addition of Gd based on a new analysis by the Daya Bay collaboration on the possibility of neutron capture with hydrogen \cite{lab17}, the relationship between the C/H ratio and attenuation length has also been studied in this work. The hope is to find a suitable low C/H ratio liquid scintillator with excellent light absorption properties to be used in the JUNO experiment.

Fig. 9 shows the C/H ratio obtained through an Element Analysis technique (CHN-O-Rapid), and attenuation length and the relationship therebetween for the LAB samples produced by Jinling. It can be observed that the C/H ratio has a non-trivial relationship with the attenuation length.

\begin{center}
\includegraphics[width=8cm]{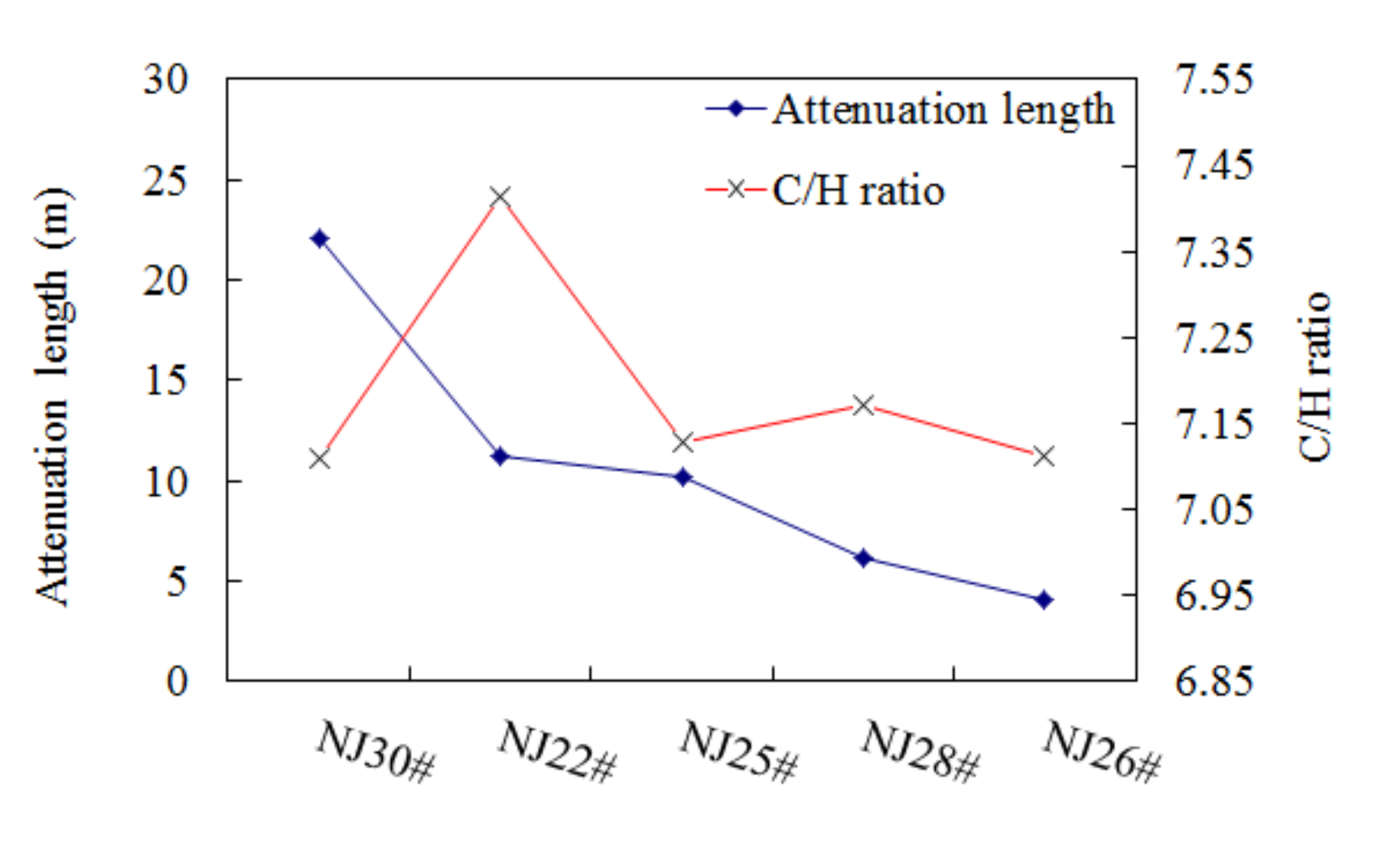}
\figcaption{\label{fig9}C/H ratio and attenuation length of the LAB samples and the relationship therebetween.}
\end{center}

\section{Discussion}

In the attenuation length measurement experiment, it can be observed from Fig. 6 that a high light intensity can lead to deviations from Equation (1). This suggests that the PMT might have effectively lost its linearity property at this intensity magnitude.

In Fig. 8 whereby the liquid height was reduced in small steps, in this case, 0.02 m, until the next desired height to be measured for its corresponding attenuated light intensity, the data for NJ32\# is not well-fitted with an exponential curve that is consistent with Eqn. (2), but nonetheless, shows the largest attenuation length that has been measured from our samples yet. In comparison, we observe a better agreement between the data and Eqn. (2) in Fig. 6 and 7 whereby the liquid height was reduced in one full step. Investigation into the cause of this is still in progress.

Overall, the concentration of the impurities in the LAB samples was sufficiently low. Even if any influence thereof on the attenuation length of LAB by the inorganic impurities may be currently negligible, the case for organic impurities playing an important role in reducing the attenuation length of LAB samples still has to be investigated deeply. In particular, we will be interested in identifying the functional groups among the organic impurities with the largest impact on the attenuation length.

As have been aforementioned, the C/H ratio has a non-trivial relationship with the attenuation length for the LAB samples that we have measured. Our prior expectation was that the C/H ratios of the samples are about the same level within uncertainties assuming that the organic impurities influences therein are negligible. Clearly, this is not the case from the results as shown in Fig. 9. Since samples NJ22\# and NJ28\# have a much higher C/H ratio compared to the other samples, it is possible that both these samples have a higher contamination of organic impurities.

\section{Summary}

In this work, the light absorption properties of different LAB samples have been studied. The attenuation length of the samples has been quantified with measurements performed in a clean dark room under a less than 0.5 $^{\circ}$C ambient temperature fluctuation with an acquired PMT signal stable to within 0.2 \%. Furthermore, the inorganic impurity content of the samples has been found to be negligible, and thus, we find no evidence that inorganic impurities have any significant impact on the attenuation length of the samples. The C/H ratio has been found to have a non-trivial relationship with the corresponding measured attenuation length of the samples. Further study on the light absorption properties of LAB and improvement of the measurement techniques thereof is in progress.

\vspace{3mm}

\textit{We are thankful for the helpful discussions and suggestions, including from Chen Shen-Jian, Wang Yi-Fang, Cao Jun, Heng Yue-Kun, Qian sen, Zhou Li, Ding Ya-Yun, Yu Bo-Xiang, Ning Zhe and Xia Jing-Kai. This work was supported by National 973 Project Foundation of the Ministry of Science and Technology of China (Contract No. 2013CB834300).}

\end{multicols}

\vspace{-1mm}
\centerline{\rule{80mm}{0.1pt}}
\vspace{2mm}

\begin{multicols}{2}

\end{multicols}

\clearpage


\begin{thebibliography}{90}

\vspace{3mm}

\bibitem{lab1}JUNO CDR, JUNO Collaboration. 22 December 2014.

\bibitem{lab2}Liang Zhan, Yifang Wang, Jun Cao et al. PHYSICAL REVIEW D {\bf 78}, 111103(R) (2008).

\bibitem{lab3}Liang Zhan, Yifang Wang, Jun Cao et al. PHYSICAL REVIEW D {\bf 79}, 073007 (2009).

\bibitem{lab4}Yu-Feng Li, Jun Cao, Yifang Wang et al. PHYSICAL REVIEW D {\bf 88}, 013008 (2013).

\bibitem{lab5}F. P. An, J. Z. Bai, A. B. Balantekin et al (Dayabay Collaboration). Phys. Rev. Lett. {\bf 108}, 171803.

\bibitem{lab6}Johnny Goett, James Napolitano, Minfang Yeh et al. Nuclear Instruments and Methods in Physics Research A {\bf 637} (2011) 47-52.

\bibitem{lab7}Yayun Ding, Zhiyong Zhang, Jinchang Liu et al. Nuclear Instruments and Methods in Physics Research A {\bf 584} (2008) 238-243.

\bibitem{lab8}P. Huang, P. Li, Z. Fu et al. 2010 JINST {\bf 5} P08007.

\bibitem{lab9}Pin-Wen Huang, Hui-Ying Cao, Ming Qi et al. Theor Chem Acc (2011) {\bf 129}:229-234.

\bibitem{lab10}Li Pi-Yi, Huang Pin-Wen, Fu Zai-Wei et al. NUCLEAR TECHNIQUES, Vol. {\bf 33}, No. 8 Aug. 2010.

\bibitem{lab11}Liu Jin-Chang, Li Zu-Hao, Yang Chang-Gen et al. HIGH ENERGY PHYSICS AND NUCLEAR PHYSICS, Vol. {\bf 31}, No. 1 Jan. 2007.

\bibitem{lab12}Ning Zhe, Qian Sen, Fu Zai-Wei et al. NUCLEAR TECHNIQUES, Vol.{\bf 33}, No.10 October 2010.

\bibitem{lab13}Dennis H. Oh, Mitsuru Sano, Steven C. Boxer et al. Soc. 1991, {\bf 113}, 6880-6890.

\bibitem{lab14}Mirco G. Colombo, Andreas Hauser, and Hans U. Gudel, Inorg. Chem. 1993, {\bf 32}, 3088-3092.

\bibitem{lab15}GAO Long, YU Bo-Xiang, DING Ya-Yun et al. Chinese Physics C, Vol. {\bf 37}, No. 7 (2013) 076001.

\bibitem{lab16}Wanda Beriguete, Jun Cao, Yayun Ding et al. Nuclear Instruments and Methods in Physics Research A {\bf 763} (2014) 82-88.

\bibitem{lab17}F. P. An, A. B. Balantekin, H. R. Band et al (Dayabay Collaboration). Phys. Rev. D {\bf 90}, 071101 (2014).


\end{thebibliography}
\end{document}